\documentclass[pre,twocolumn,superscriptaddress]{revtex4}
\usepackage{latexsym}

\usepackage{graphicx}
\usepackage{amssymb}
\usepackage{amsmath}

\usepackage{mathptmx}       
\usepackage{helvet}         
\usepackage{courier}        
\usepackage{amsthm,amsmath,amsfonts,amssymb,latexsym}

\newcommand{\R}{\mathbb R}

\def\be#1\ee{\begin{equation}#1\end{equation}}

\newcommand{\bq}{\begin{equation}}
\newcommand{\eq}{\end{equation}}

\def\bqa{\begin{eqnarray}}
\def\eqa{\end{eqnarray}}

\newcommand{\bd}{\begin{displaymath}}
\newcommand{\ed}{\end{displaymath}}
\newcommand{\ba}{\begin{eqnarray}}
\newcommand{\ea}{\end{eqnarray}}

\def\R{\mathbb{R}}

\begin{document}

\author{D. Maldarella}
\affiliation{Department of Mathematics \& CMCS, University of Ferrara,  via Machiavelli 35, Ferrara, Italy. }
\email[ e-mail addresses: ]{dario.maldarella@unife.it, lorenzo.pareschi@unife.it}
\author{L. Pareschi}
\affiliation{Department of Mathematics \& CMCS, University of Ferrara,  via Machiavelli 35, Ferrara, Italy. }

\date{\today}

\title{Price dynamics in financial markets: a kinetic approach}

\begin{abstract}
The use of kinetic modelling based on partial differential
equations for the dynamics of stock price formation in financial
markets is briefly reviewed. The importance of behavioral aspects
in market booms and crashes and the role of agents' heterogeneity
in emerging power laws for price distributions is emphasized and
discussed.
\end{abstract}

\maketitle

\section{Introduction}
The recent events of the 2008 world's financial crisis and its
uncontrolled effect propagated among the global economic system,
has produced a deep rethink of some paradigm and fundamentals in
economic modelling of financial markets \cite{Ad}. A big amount of
efforts has been done in the understanding of stock's price
dynamics, but also in the attempt to derive useful models for the
risk estimation or price prediction. Nevertheless the need to find
a compromise between the extraordinary complexity of the systems
and the request of quite simplified models from which some basic
information can be derived, represents a big challenge, and it is
one of the main difficulties one have to deal with, in the
construction of models.

Any reasonable model need to rely on some fundamental hypotheses
and to rest on a theoretical framework, which should be able to
provide some basic and universal principles, this is the way all
the models arising from the physical world are build up.
Unfortunately, this is not an easy task when we deal with economic
and financial systems. Looking at stock market in particular, it
is not obvious to understand which are the fundamental dynamics to
be considered and which aspects can be neglected in order to
derive the basic issues.

One of the most classical approach has been to consider the {\em
efficient market hypothesis} \cite{EF, EF1}. It relies on the
belief that securities markets are extremely efficient in
reflecting information about individual stocks and about the stock
market as a whole. When information arises, the news spread very
quickly and are incorporated into the prices of securities without
delay. Thus, neither technical analysis, which is the study of
past stock prices in an attempt to predict future prices, nor even
fundamental analysis, which is the analysis of financial
information such as company earnings, asset values, etc., to help
investors select undervalued stocks, would enable an investor to
achieve returns greater than those that could be obtained by
holding a randomly selected portfolio of individual stocks with
comparable risk.

The efficient market hypothesis is associated with the idea of a
{\em random walk} \cite{Ba, CP, EF1}, which is widely used in the
finance literature to characterize a price series where all
subsequent price changes represent random departures from previous
prices. The logic of the random walk idea is that if any
information is immediately reflected in stock prices, then
tomorrow's price change will reflect only tomorrow's news and will
be independent of the price changes today. Thus, resulting price
changes must be unpredictable and random.

Strongly linked to the market efficiency hypothesis, is the
assumption of {\em rational behavior} among the traders.
Rationality of traders can be basically reassumed in two main
features. First, when they receive new information, agents update
their beliefs by evaluating the probability of hypotheses
accordingly to Bayes' law. Second, given their beliefs, agents
make choices that are completely rational, in the sense that they
arise from an optimization process of opportune subjective utility
functions.

This traditional framework is appealingly simple, and it would be
very satisfying if its predictions were confirmed in the data.
Unfortunately, after years of efforts, it has become clear that
basic facts about the stock market, the average returns and
individual trading behavior are not easily understood in this
framework \cite{RL}.

By the beginning of the twenty-first century, the intellectual
dominance of the efficient market hypothesis had become far less
universal. Many financial economists and statisticians began to
believe that stock prices are at least partially predictable. A
new breed of economists emphasized psychological and behavioral
elements of stock-price determination. The {\em behavioral
finance} approach has emerged in response to the difficulties
faced by the traditional paradigm \cite{EF2, SH, SH1}. It relies
in the fact that some financial phenomena can be better understood
using models in which some agents are not fully rational. In some
behavioral finance models, agents fail to update their beliefs
correctly. In other models, agents apply Bayes' law properly but
make choices that are questionable, in the sense that they are
incompatible with the optimization of suitable utility functions.

A strong impact in the field of behavioral finance has been given
by the introduction of the {\em prospect theory} by Kahneman and
Tversky \cite{KT, KT1}. They present a critique of expected
utility theory as a descriptive model of decision making under
risk and develop an alternative model. Under prospect theory,
value is assigned to gains and losses rather than to final assets
and probabilities are replaced by decision weights. The theory
which they confirmed by experiments predicts a distinctive
fourfold pattern of risk attitudes: risk aversion for gains of
moderate to high probability and losses of low probability, and
risk seeking for gains of low probability and losses of moderate
to high probability. Further development in this direction were
done by De Bondt and Thaler \cite{WBT} who effectively form the
start of what has become known as behavioral finance. They
discovered that people systematically overreacting to unexpected
and dramatic news events results in substantial weak-form
inefficiencies in the stock market.

Recently, {\em agent based modelling} methods have given an
important contribute and provided a huge quantity of numerical
simulations \cite{KM, LLS, Lux, LM, LM1}. The idea is to produce a
big mass of artificial data and to observe how they can fit with
empirical observations. This approach is now also supported by the
availability of many recorded empirical data \cite{SZSL}. The aim
of the construction of such microscopic models of financial
markets is to reproduce the observed statistical features of
market movements (e.g. fat tailed return distributions, clustered
volatility, cycles, crashes) by employing highly simplified models
with large numbers of agents. Microscopic models of financial
markets are highly idealized as compared to what they are meant to
model \cite{SZSL}. The relevant part of physics that is used to
build such models of financial markets consists in methods from
statistical mechanics. This attempt by physicists to map out the
statistical properties of financial markets considered as complex
systems is usually referred to as econophysics \cite{MS, GGPS,
VT}.

The need to recover mathematical models which can display such
scaling properties, but also capable to deal with systems of many
interacting agents and to take into account the effects of
collective endogenous dynamics, put the question on the choices of
the most appropriate mathematical framework to use. In fact,
besides numerical simulations, it is of paramount importance to
have a rigorous mathematical theory which permits to identify the
essential features in the modelling originating the stylized
facts. The classical framework of stochastic differential
equations which played a major rule in financial mathematics seems
inadequate to describe the dynamics of such systems of interacting
agents and their emerging collective behavior.

In the last years a new approach based on the use of {\em kinetic
and mean field models} and related mathematical tools has appeared
in the mathematics and physics community \cite{BM, CCS, CPP, CPT,
DMTb, IKR, LLi}. We refer to \cite{NPT} for a recent review on the
use of microscopic and kinetic modelling in socio-economic
sciences. Kinetic theory was introduced in order to give a
statistical description of systems with many interacting
particles. Rarefied gases can be thought as a paradigm of such
complex systems, in which particles are described by random
variables which represents their physical states, like position
and velocity. A Boltzmann equation then prescribes the time
evolution for the particles density probability function
\cite{Cer}. This seems to fit very well with the necessity to
prescribe how the trading agents interacting in a stock market are
leaded to form their expectations and revaluate their choices on
the basis of the influence placed on the neighbor agents' behavior
rather than the flux of news coming from some fundamental analysis
or direct observations of the market dynamic. The kinetic approach
reveals particularly powerful when from some simple local
interaction rules some global features for the whole system has to
be derived, but also in the study of asymptotic regimes and
universal behaviors described by Fokker-Planck equations.

Here we briefly review some recent advances in this direction
concerned with the kinetic modelling of the price dynamics in a
simple stock market where two types of agents interact \cite{MP}.
Other kinetic and mean field approaches to price formation have
been considered in \cite{CPP, LLi}. Most of what we will present
here has been inspired by the work of Lux and Marchesi \cite{LM,
LM1, Lux} on microscopic models for the stock market. Quite
remarkably, however, behavioral features are taken into account in
our model. In spite of its structural simplicity the kinetic model
is able to reproduce many stylized facts such as lognormal and
power laws price profiles and the appearance of market booms,
crashes and cycles. As it is shown, non rational behavioral
aspects and agents' heterogeneity are essential components in the
model to achieve such behaviors.

\section{Kinetic modelling for price formation}

\subsection{Opinion modelling}
The collective behavior of a system of trading agents can be
described by introducing a state variable $y\in[-1,1]$ and the
relative density probability function $f(y)$ which, for each
agent, represent respectively the propensity to invest and the
probability to be in such a state. Positive values of $y$
represent potential buyers, while negative values characterize
potential  sellers, close to $y=0$ we have undecided agents.
Clearly
\be
\rho(t)=\int_{-1}^1 f(y,t)\,dy,
\ee
represents the number density. Moreover we define the mean
investment propensity \be Y(t)= \frac1{\rho(t)}\int_{-1}^1
f(y,t)y\,dy.\ee Traders are allowed to compare their strategies
and to revaluate them on the basis of a compromise opinion
dynamic. This is done by assigning simple binary interaction
rules, where, if the pair $(y,y_*)$ and $(y',y'_*)$ represent
respectively the pre-interaction and post-interaction opinions, we
have
\begin{eqnarray}
\nonumber
y'&=&(1-\alpha_1 H(y))y+\alpha_1 H(y){y}_*+ D(y)\eta,\\[-.1cm]
\\[-.1cm]
\nonumber
y_*'&=&(1-\alpha_1 H(y_*))y_*+\alpha_1 H(y_*){y} +
D(y_*)\eta_*.
\end{eqnarray}
Here $\alpha_1\in [0,1]$ measures the importance the individuals
place on others opinions in forming expectations about future
price changes. The random variables $\eta$ and $\eta_*$ are
assumed distributed accordingly to $\Theta(\eta)$ with zero mean
and variance $\sigma^2$ and measure individual deviations to the
average behavior. The functions $H(y)$ and $D(y)$ characterize
respectively, herding and diffusive behavior. Simple examples of
{herding function} and {diffusion function} are given by
\[
H(y)=a+b(1-|y|),\qquad D(y)=(1-y^2)^{\gamma},
\]
with $0\leq a+b\leq 1$, $a\geq 0, b>0$, $\gamma > 0$. A kinetic
model for opinion formation based on such interactions was
recently introduced by Toscani \cite{TG}.

\subsection{Market influence}

The traders are also influenced by the dynamics of stock market's
price, so a coupling with the price dynamic has to be considered.
With the same kinetic setting we define the probability density
$V(s,t)$ of a given price $s$ at time $t$. The market price $S(t)$
is then defined as the mean value
\be
S(t)=\int_0^\infty V(s,t) s\,ds.
\ee
Price changes are modeled as endogenous responses of the market to
imbalances between demand and supply characterized by the mean
investment propensity accordingly to the following price
adjustment
\be
s' = s + \beta \rho Y(t)s + \eta s,
\ee
where $\beta>0$ represents the price speed evaluation and $\eta$
is a random variable with zero mean and variance $\zeta^2$
distributed accordingly to $\Psi(\eta)$.

\begin{figure}[htb]
\includegraphics[scale=0.4]{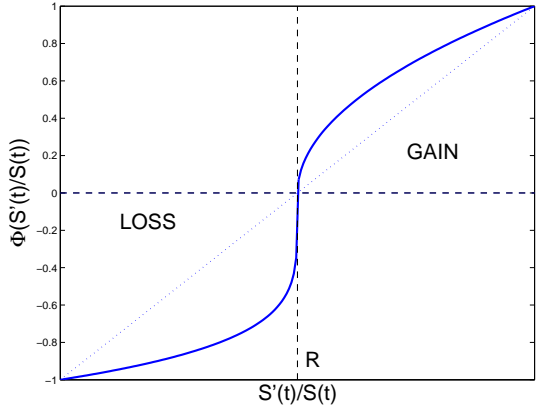}
\caption{An hypothetical value function. The reference point $R$
is the value of $S'/S$ such that $\Phi(R)=0$. The value function
decision makers use to assess the possible shifts away from the
reference point is concave in the domain of gains and convex in
the domain of losses.}
\end{figure}

To take into account the influence of the price in the mechanism
of opinion formation of traders, we introduce a normalized value
function $\Phi=\Phi(\dot{S}(t)/{S(t)})$ in $[-1,1]$ in the sense
of Kahneman and Tversky \cite{KT, KT1} that models the reaction of
individuals towards potential gain and losses in the market. Thus
we reformulate the binary interaction rules in the following way
\begin{eqnarray}
\nonumber
y'&=&(1-\alpha_1 H(y)-\alpha_2)y+\alpha_1 H(y){y}_*+\alpha_2\Phi+
D(y)\eta,\\[-.1cm]
\\[-.1cm]
\nonumber
y_*'&=&(1-\alpha_1 H(y_*)-\alpha_2)y_*+\alpha_1
H(y_*){y}+\alpha_2\Phi+D(y_*)\eta_*.
\end{eqnarray}
Here $\alpha_1\in [0,1]$ and $\alpha_2 \in [0,1]$, with $\alpha_1
+ \alpha_2 \leq 1$, measure the importance the individuals place
on others opinions and actual price trend in forming expectations
about future price changes. This permits to introduce behavioral
aspects in the market dynamic and to take into account the
influence of psychology and emotivity on the behavior of the
trading agents.

Note that agents influence the price through their mean propensity
to invest $Y(t)$ and at the same time the price trend influences
their mean propensity through the value function $\Phi$. Thus,
except for the particular shape of the value function, if the mean
propensity is initially (sufficiently) positive then it will
continue to grow together with the price and the opposite occurs
if it is initially (sufficiently) negative. The market goes
towards a {boom} (exponential grow of the price) or a {crash}
(exponential decay of the price).

\subsection{Lognormal behavior}

A set of Boltzmann equations for the evolution of the unknown
densities $f(y,t)$ and $V(s,t)$ can be obtained using the standard
tools of kinetic theory \cite{Cer}. Such system reads
\begin{eqnarray}
\nonumber
\frac{\partial f}{\partial t} &=& Q(f,f),\\[-.3cm]
\\
\label{eq:kinetic}
\nonumber \frac{\partial V}{\partial t} &=&
L(V),
\end{eqnarray}
where the quadratic operator $Q$ and the linear operator $L$ can
be conveniently written in weak form as
\begin{eqnarray*}
&&\int_{-1}^{1}Q(f,f)\varphi(y)\,dy=\\&&\int_{[-1,1]^2}\int_{\R^2}
B(y,y_*)
f(y)f(y_*)(\varphi(y')-\varphi(y))d\eta\,d\eta_*\,dy_*\,dy,\\
&&\int_{0}^{\infty}L(V)\varphi(s)\,ds=\int_{0}^{\infty}\int_{\R}
b(s) V(s)(\varphi(s')-\varphi(s))d\eta\,ds.
\end{eqnarray*}
In the above equations $\varphi$ is a test function and the
transition rates have the form
\begin{eqnarray*}
B(y,y_*)&=&\Theta(\eta)\Theta(\eta_*)\chi(|y'|\leq
1)\chi(|y_*'|\leq 1),\\ b(s)&=&\Psi(\eta)\chi(s'\geq 0),
\end{eqnarray*}
with $\chi(\cdot)$ the indicator function.

A simplified Fokker-Planck model which preserves the main features
of the original Boltzmann model is obtained under a suitable
scaling of the system. In such scaling all agents interact
simultaneously with very small variations of their investment
propensity (see \cite{MP} for details). This allows us to recover
the following Fokker-Plank system \be \left\{
\begin{array}{lcl}
\displaystyle\frac{\partial {f}}{\partial t} &+&\displaystyle
\frac{\partial}{\partial y}\left[ \left(\rho{\alpha_{1}}
H(y)({Y}-y) + \rho{\alpha_{2}} \left(\Phi
-y\right)\right){f}\right]\\ \displaystyle &=& \displaystyle
\frac{\sigma^2\rho}{2}\frac{\partial^{2}}{\partial
y^{2}}[{D}^{2}(y){f}],\\
\displaystyle\frac{\partial V}{\partial t}
&+&\displaystyle\frac{\partial}{\partial s}\left({\beta}\rho{Y} s
{V}\right) = \frac{\zeta^2}{2}\frac{\partial^{2}}{\partial
s^{2}}\left( s^{2}{V}\right),
\end{array}
\right. \label{eq:fps} \ee where we kept the original notations
for all the scaled quantities.

The above equation for the price admits the self similar lognormal
solution \cite{CPP, MP} \be {V}(s,t) = \frac{1}{s(2\log(Z(t)^2)
\pi)^{\frac{1}{2}}}\exp\left( -\frac{( \log(s
Z(t))^{2}}{2\log(Z(t)^2)}\right), \label{eq:logp} \ee where
$Z(t)=\sqrt{E(t)}/S(t)$, and $E(t)$ satisfies the differential
equation
\[
\frac{dE}{dt}=(2{\beta}Y +\zeta^2) E.
\]

\section{Playing a different strategy}

We consider now in the stock market the presence of traders who
deviate their strategy from the mass. We introduce trading agents
who rely in a fundamental value for the traded security. They are
buyer while the price is below  the fundamental value and seller
while the price is above. Expected gains or losses are then
evaluated from deviations of the actual market price and just
realized only wether or not the price will revert towards the
fundamental value. Such agents are not influenced by other agents'
opinions.

The microscopic interactions rules for the price formation now reads
\be
s' = s + \beta (\rho Y(t)s + \rho_{F}\gamma(S_{F}-s)) + \eta s,
\ee
where $S_F$ represent the fundamental value of the price, $\rho_f$
is the number density for such trading agents performing a
different strategy, while  $\gamma$ is the reaction strength to
deviations from the fundamental value. If we are interested in
steady states we can ignore the possibility of a strategy exchange
between traders and the resulting kinetic system has the same
structure (\ref{eq:kinetic}). We refer to \cite{MP} for a complete
treatment of a model including strategy exchanges.

\subsection{Equilibrium states}
The system of equations (\ref{eq:kinetic}) in the simplified case
of $H$ constant admits the following possible macroscopic
equilibrium configurations \cite{MP}
\begin{itemize}
\item[$(i)$] $\quad \rho_F\neq 0,\quad S=S_F,\quad Y=0,\quad \Phi(0)=0$,
\item[$(ii)$] $\quad \rho_F=0,\quad Y=0, \quad \Phi(0)=0, \quad
S$\,\,{arbitrary},
\item[$(iii)$] $\quad \rho_F=0,\quad Y=Y_*$,\, with\, $Y_*=\Phi(\beta t_C Y_*), \quad
S=0$,
\end{itemize}
where only configuration $(i)$ takes into account the presence of
both types of traders. Note, however, that if the reference point
for the value function is different from zero, namely $\Phi(0)\neq
0$, configuration $(i)$ and $(ii)$ are not possible. This is in
good agreement with the fact that an emotional perception of the
market acts as a source of instability for the market itself. In
contrast configuration $(iii)$, corresponding to a market crash,
can be achieved also for $\Phi(0)\neq 0$.

\subsection{Emergence of power laws}
The presence of fundamentalists leads to the following
Fokker-Plank equation for the probability density function $V$
\begin{eqnarray}
\frac{\partial V}{\partial t} +\frac{\partial}{\partial
s}\left[{\beta}\left(\rho {Y} s +
\rho_F\gamma({S}_{F}-s)\right){V}\right] =
\frac{\zeta^2}{2}\frac{\partial^{2}}{\partial s^{2}}\left(
s^{2}{V}\right).\\[-.1cm]
\nonumber
\end{eqnarray}
If we consider the equilibrium configuration $(i)$ a steady state
for the Fokker-Planck equation can be computed in the form of a
Gamma distribution \cite{BM, CPT, MP}
\begin{equation}
{V}^{\infty}(s) =
C_{1}(\mu)\frac{1}{s^{1+\mu}}e^{-\frac{(\mu-1)S_F}{s}},
\end{equation}
where $\mu=1+2\rho_F\gamma /\zeta^2$ and
$C_1(\mu)=((\mu-1)S_F)^\mu/\Gamma(\mu)$ with $\Gamma(\cdot)$ being
the usual Gamma function. Therefore the price distribution
exhibits a Pareto tail behavior.

\end{document}